\documentclass[]{article}
\usepackage{latexsym}
\usepackage{epsfig}

\Large

\long\def\@makecaption#1#2{%
  \vskip\abovecaptionskip
  \sbox\@tempboxa{\pushziti\small\rm\songti\zihao{-5}#1: #2\popziti}%
  \ifdim \wd\@tempboxa >\hsize
    \box\@tempboxa\par
  \else
    \global \@minipagefalse
    \hbox to\hsize{\hfil\box\@tempboxa\hfil}%
  \fi
  \vskip\belowcaptionskip}

\begin{document}

\centerline{\Large{\bf Nonstandard Picture of Turbulence (The
Second Revised)}}

$${}$$
\centerline{Feng \quad Wu }\centerline{\footnote{klfjgjg}}
\centerline{\it Department of Mechanics
and Mechanical Engineering,}

\centerline{\it University of Science and Technology of China, Hefei 230026, China}
$${}$$
\setlength{\baselineskip}{25pt}

\noindent \small{The nonstandard picture of a turbulent field is
presented in the article. By the concepts of nonstandard
mathematics, the picture describes the hierarchical structure of
turbulence and shows the mechanism of fluctuation appearing in a
turbulent field. The uncertainty of measurement is pointed out.
And the fundamental equations of turbulence are given too. It is
natural in this picture that the reasonable closure methods can be
obtained and seem to be precise.}

$${}$$

\large
\setlength{\baselineskip}{25pt}
\section{Some Concepts of Nonstandard Analysis}\indent
\indent The nonstandard analysis theory of turbulence will be
presented in this paper. The nonstandard analysis means the
mathematic fundament of the nonstandard picture of turbulence.
Therefore, some concepts of nonstandard analysis should be
introduced in the first place.

On the basis of  mathematic logic, A.Robinson \cite{lgc} proved in
the sixties of the last century that a real number system $\it R$
can be expanded into the hyperreal number system $\it R^{*}$.
Besides all real numbers (i.e., standard numbers), $\it R^{*}$
contains also hyperreal numbers (i.e., nonstandard numbers). The
infinitesimal $\varepsilon$ and infinite $L$ are elementary
nonstandard numbers. There are other nonstandard numbers, for
example, $\xi \pm\varepsilon, \xi\varepsilon, \xi L, \xi\pm L,
\varepsilon\varepsilon, LL,$
 etc. ($\xi$ stands for any real number).

It is known that real numbers one-to-one correspond  to the points
on a real line, but on the real line there are no such points
which correspond to nonstandard numbers. In other words
nonstandard numbers can not be indicated by the points on a real
line. The concept of monad needs to be presented to indicate the
nonstandard numbers. For every real number $\xi$, there exists a
monad that is composed of the numbers infinitely close to $\xi$
and $\xi$ belongs also to the monad. In fact, there exist infinite
monads containing the real number $\xi$. The size of all these
monads is infinitesimal $\varepsilon$. The numbers in this monad,
except $\xi$, are all nonstandard numbers of which $\xi$ is the
standard part.

So any point on a real line is expanded into a monad.  The real
number, represented by the point on the real line, is  the
standard part of the nonstandard numbers contained in the monad.
And a real line, formed by the points which indicate only real
numbers,is expanded into a hyperreal line composed of monads. The
set of the standard parts of all these monads composes the real
line mentioned above. Every monad on a hyperreal line is also
called a point (i.e.,standard point). Clearly every such standard
point is not the absolute point which has the length of absolute
zero, but the monad whose length is infinitesimal $\varepsilon$.

Every monad must contain only one real number. If a monad contains
the real number $\xi$, we call the monad as $\xi$-monad.

The physical world is in a hierarchical structure. This
hierarchical structure can be described by the concepts of
nonstandard mathematics. For example, there are three levels:
level $\alpha$ (with infinitesimal $\varepsilon$, infinite
$\varepsilon^{-1}$), level $\beta$ ( with infinitesimal
$\varepsilon^{3}$, infinite $\varepsilon$) and level $\gamma$ (
with infinitesimal $\varepsilon^{-1}$, infinite
$\varepsilon^{-3}$). Any monad on a hyperreal line in level
$\alpha$ is also a whole hyperreal line in level $\beta$. However,
the whole hyperreal line in level $\alpha$ is only a monad of the
hyperreal axis in higher level $\gamma$.  From the viewpoint of
hierarchical structure, the meanings of infinite and infinitesimal
are not fixed. It is understandable in physics. Physicists, in
fact, can even determine how large infinite and infinitesimal are
in a specific physical problem, and often take a certain number as
infinite or infinitesimal. In physics any number (not absolute
zero) is taken as infinite, infinitesimal or finite in various
specific condition.

The coordinate systems in two levels will be set up for the
convenience of description. Let space coordinates be
$(x_{1},x_{2},x_{3})$, time $t$ in level $\alpha$ and space
$(x^{\prime}_{1},x^{\prime}_{2},x^{\prime}_{3})$, time
$t^{\prime}$ in level $\beta$, the range of values of
$(x_{1}^{\prime}, x_{2}^{\prime},x_{3}^{\prime},t^{\prime})$ be
$[0,L_{1}),[0,L_{2}),[0,L_{3})$ and $[0,T)$ respectively. Here
$L_{1},L_{2},L_{3}$ and $T$ are the infinites of space and time,
respectively, in level $\alpha$. For any time-space point
$(\vec{x},t)$ in level $\alpha$, the monad of this point is
expressed by $(\vec{x},t)$ and the dimensions of the monad are
$(\varepsilon_{1},\varepsilon_{2},\varepsilon_{3},
\varepsilon_{t})$. Here
$\varepsilon_{1},\varepsilon_{2},\varepsilon_{3}, \varepsilon_{t}$
are infinitesimals of space and time, respectively, in level
$\alpha$. Moreover, assume that:

\begin{equation}
\varepsilon_{1}L_{1}=1,\quad\varepsilon_{2}L_{2}=1,\quad\varepsilon_{3}
L_{3}=1,\quad\varepsilon_{t}T=1
\end{equation}

\noindent The space-distance between any two points

$$\Delta l=\sqrt{(\Delta x_{1})^{2}+(\Delta
 x_{2})^{2}+(\Delta x_{3})^{3}}$$

\noindent in level $\alpha$, and

$$\Delta l^{\prime}=\sqrt{(\Delta x^{\prime}_{1})^{2}+(\Delta x^{\prime}_{2})^{2}
+(\Delta x^{\prime}_{3})^{2}}$$

\noindent in level $\beta$. The actual length (i.e., the length
observed from the angle of level $\alpha$) of $\Delta l^{\prime}$
is:

  \begin{equation}
  \sqrt{(\Delta x^{\prime}_{1} \varepsilon^{2}_{1})^{2}+
  (\Delta x^{\prime}_{2} \varepsilon^{2}_{2})^{2}+(\Delta x^{\prime}_{3}
   \varepsilon^{2}_{3})^{2}}
  \end{equation}

\noindent If $\varepsilon_{1}=\varepsilon_{2}=\varepsilon_{3}
=\varepsilon$  the actual length of $\Delta l^{\prime}$ is:

   \begin{equation}
   \Delta l^{\prime} \varepsilon^{2}
   \end{equation}

\noindent Similarly, the actual value (i.e., the value observed
from level $t$) of $\Delta t^{\prime}$ is:

   \begin{equation}
   \Delta t^{\prime}\varepsilon_{t}^{2}
   \end{equation}

\noindent And the actual lengths of the whole space and time  axes
in level $\beta$ are:

   \begin{equation}
   L_{1}\varepsilon_{1}^{2}=\varepsilon_{1},\quad L_{2}\varepsilon
   _{2}^{2}=\varepsilon_{2},\quad L_{3}\varepsilon_{3}^{2}=\varepsilon_{3},
   \quad T\varepsilon_{t}^{2}=\varepsilon_{t}
   \end{equation}

\noindent They are just the dimensions of a monad in level $\alpha$.

Now the limit of tending to zero ($\Delta x_{i}\rightarrow 0$) in
level $\alpha$ is not tending to absolute zero but infinitesmal
$\varepsilon$. Here $\Delta x_{i}$ can not go into the inner of
the zero monad. In the inner of the zero monad in level $\alpha$
exists $\Delta x^{\prime}_{i}$ instead of $\Delta x_{i}$. In other
words, the process of tending to zero does not mean tending to
absolute zero but the dimension of a monad in the same level. The
standard part of any number $\omega$ is denoted by $St\omega$. To
show clearly the meanings of $\Delta x_{i}\rightarrow 0$ and
$\Delta t\rightarrow 0$ mentioned above, we do not use the
notation $\Delta x_{i}\rightarrow 0$ and $\Delta t\rightarrow 0$.
Instead, we use

\begin{equation}
St\Delta x_{i}\rightarrow 0,\quad\quad St\Delta t\rightarrow 0
\end{equation}

\noindent i.e.,
\begin{equation}
    \lim_{St\Delta x_{1}\rightarrow0} \Delta x_{1}=\varepsilon_{1},\lim_{St\Delta x_{2}
    \rightarrow 0}\Delta x_{2}=\varepsilon_{2},\lim_{St\Delta x_{3}\rightarrow0}\Delta x
   _{3}=\varepsilon_{3},\quad\lim_{St\Delta t\rightarrow0}\Delta t=\varepsilon_{t}
\end{equation}

\noindent In other words, the definitions of ($\Delta
x_{i}\rightarrow \varepsilon_{i}$) and ($\Delta
t\rightarrow\varepsilon_{t}$) are, respectively, ($St\Delta
x_{i}\rightarrow 0$) and ($St\Delta t\rightarrow 0$). Similarly,
the limit of tending to infinite does not mean tending to absolute
infinite but the nonstandard number $L$, i.e.,
   \begin{equation}
   \lim_{St\Delta x_{1}\rightarrow\infty}\Delta x_{1}=L_{1}, \lim_{St\Delta x_{2}
   \rightarrow\infty}\Delta x_{2}=L_{2}, \lim_{St\Delta x_{3}\rightarrow\infty}
   \Delta x_{3}=L_{3}, \quad\lim_{St\Delta t\rightarrow\infty}\Delta t=T
   \end{equation}

Moreover, in standard case, a function is expressed by
$f(x_{1},x_{2},x_{3},t)$ which represents the function value at
the standard point $(x_{1},x_{2},x_{3},t)$. However, in the
nonstandard analysis, a point $(x_{1},x_{2},x_{3},t)$ is a monad
within which there are still infinite nonstandard points
$(x_{1}^{\prime}, x_{2}^{\prime},x_{3}^{\prime},t^{\prime})$. Then
a function should be expressed by
$f(x_{1},x_{2},x_{3},t,x_{1}^{\prime},x_{2}^{\prime},
x_{3}^{\prime},t^{\prime})$ which is the function value at the
nonstandard point
$(x_{1}^{\prime},x_{2}^{\prime},x_{3}^{\prime},t^{\prime})$
contained in the monad $(x_{1},x_{2},x_{3},t)$.  Still we define
the partial derivatives in form:

\begin{equation}
   \frac{\partial f}{\partial x_{i}^{\prime}}=\lim_{St\Delta x^{\prime}_{i}\rightarrow 0}
   \frac{f(x_{i}^{\prime}+\Delta x^{\prime}_{i})-
   f(x_{i}^{\prime})}{\Delta x^{\prime}_{i}},\quad\frac{\partial f}{\partial t^{\prime}}
   =\lim_{St\Delta t^{\prime}\rightarrow 0}\frac{f(t^{\prime}+\Delta t^{\prime})
   -f(t^{\prime})}{\Delta t^{\prime}}
\end{equation}

\begin{equation}
   \frac{\partial f}{\partial x_{i}}=\lim_{St\Delta x_{i}\rightarrow 0}
   \frac{f(x_{i}+\Delta x_{i})
   -f(x_{i})}{\Delta x_{i}},\quad\frac{\partial f}{\partial t}=\lim_{St\Delta t\rightarrow 0}
   \frac{f(t+\Delta t)-f(t)}{\Delta t}
\end{equation}

   $$(i=1,2,3)$$

\noindent Definition (9)-(10) is only a definition without the
ordinary meaning of derivative in the standard case. The
difference of $f(x_{i}+\Delta x_{i})-f(x_{i})$ in (10) is the
abbreviation of $f(x_{i}+\Delta
x_{i},x_{j},x_{k},t,x_{i}^{\prime},x_{j}^{\prime},
x_{k}^{\prime},t^{\prime})-f(x_{i},x_{j},x_{k},t,x_{i}^{\prime},x_{j}^{\prime}
,x_{k}^{\prime},t^{\prime})$, the increment of the function value
between two monads of the points $(x_{i}+\Delta x_{i})$ and
$(x_{i})$. The distance between the points
$$\lim_{St\Delta x_{i}\rightarrow 0} (x_{i}+\Delta x_{i},x^{\prime}_{i})$$
and $$(x_{i},x_{i}^{\prime})$$ is infinite from the angle of level
$\beta$.  Therefore, $\frac{\partial f}{\partial x_{i}}$ and
$\frac{\partial f}{\partial t}$ in (10) have the meanings of an
average. If one insists on the concept of ordinary derivative in
standard case, $\frac{\partial f}{\partial x_{i}}$ and
$\frac{\partial f}{\partial t}$ in (10) have no sense.

\section{Nonstandard Model Of Turbulence}\indent
\indent On the basis of the above-mentioned concepts of nonstandard analysis
a nonstandard description of turbulent field can be presented as follows.

\begin{quote}
    \it{Assumption 1: Global turbulent field is composed of standard
   points, and every standard point is yet a monad. Each monad possesses
   the internal structure, namely a monad is also composed of infinite
   nonstandard points (so called interior points of the monad)}.
\end{quote}

It is well known that a space is formed from the points. In fluid
mechanics, fluid mechanists always take  the fluid element (fluid
particle) as point. In fact, a fluid particle has the volume not
to be absolute zero but to be a micro volume, which is taken as
infinitesimal from macroscopic viewpoint and infinite from
microscope. People generally think of the whole fluid particle as
uniform. Namely the fluid particle is taken to be equal to a point
in reality. Now, in the case of turbulence, the motion of fluid
varies so fast that a fluid particle can not be uniform as a whole
but becomes a monad. From now on in this paper, the meanings of a
fluid particle, a monad on a whole and a standard point are
identical with each other and so are the meanings of a fluid
particle in the lower level, an interior point of a monad and a
nonstandard point.

Since the point of the turbulent field becomes a monad with
internal structure, the motion features of different interior
points of the monad are different. So it is obvious that there is
a flow or flow field in the inner of the monad. Hence, a fluid
particle (monad) is made up of nonstandard points. The volume of
these nonstandard points is not absolute zero, and these
nonstandard points, in fact, are the fluid particles in the lower
level. Now the particles in the lower level are thought to be
uniform. They are points (nonstandard points) actually. Moreover,
a lot of  fluid molecules are included still in the fluid particle
of the lower level.

Therefore, two kinds of flow fields exist in turbulence. One is
the global turbulent field and the other is the flow field in a
monad ( called monad field in this paper). They are flow fields in
two different levels. Now the global turbulent field is not
composed of the points in which there are no structures, but of
monad fields. So there are three levels in turbulent field:
molecular movement, monad field and global field.

\begin{quote}
     \it{Assumption 2: The flows in monad fields are controlled by the
      Navier-Stokes equations.}.
\end{quote}

Then the motion-equations of the flows in monad fields
can be written as follows (for incompressible fluid, unsteadiness
and three dimensions):

\begin{equation}
    \frac{\partial U_{i}}{(\varepsilon^{2}\partial x^{\prime})_{i}}=0
\end{equation}

\begin{equation}
    \frac{\partial U_{i}}{\varepsilon_{t}^{2}\partial t^{\prime}}+\frac
    {\partial U_{i}U_{j}}{(\varepsilon^{2}\partial x^{\prime})_{j}}=
    \frac{1}{\rho}\frac{\partial \sigma_{ij}}{(\varepsilon^{2}\partial x^{\prime})_{j}}
\end{equation}

\noindent Here $U_{i}$ is velocity component in $i$-direction,
$\rho$ density of fluid, $\sigma_{ij}$ stress tensor and
$\sigma_{ij}=\sigma_{ji}$. The independent variables of these
functions in (11)-(12) are $(x_{1},x_{2},x_{3},t,x_{1}^{\prime},
x_{2}^{\prime},x_{3}^{\prime},t^{\prime})$. $\varepsilon_{t},
\varepsilon_{1}, \varepsilon_{2}, \varepsilon_{3}$ are
infinitesimals (i.e., the linear dimensions of a monad) of time
$t$, space $x_{1},x_{2}$ and $x_{3}$, respectively. Let
$\varepsilon_{t}=\varepsilon_{1}=\varepsilon_{2}=\varepsilon_{3}=\varepsilon$
here and later, then the governing equations becomes:

\begin{equation}
   \frac{\partial U_{i}}{\partial x_{i}^{\prime}}=0
\end{equation}

\begin{equation}
    \frac{\partial U_{i}}
   {\partial t^{\prime}}+\frac{\partial U_{i}U_{j}}{\partial x_{j}^{\prime}}=
   \frac{1}{\rho}\frac{\partial \sigma_{ij}}{\partial x_{j}^{\prime}}
\end{equation}

\noindent The governing equations (11)-(12) show the internal structure of a monad.

For actual turbulence the possible linear dimensions of flow
fields in two levels could be estimated roughly as follows:

\hspace{0cm}Global turbulent field:\quad infinite\hspace{2.0cm}$L=\varepsilon^{-1}\sim 10^{1.5}cm$

\hspace{3.9cm}             \quad infinitesimal\hspace{1.0cm}$\varepsilon\sim 10^{-1.5}-10^{-2} cm$

\hspace{8.0cm}                                                (linear dimension of a monad)

\hspace{0cm}Monad field:\hspace{2.2cm}infinite\hspace{2.0cm}$\varepsilon\sim 10^{-1.5}-10^{-2} cm$

\hspace{4.3cm}           infinitesimal\hspace{0.9cm} $\varepsilon^{3} \sim 10^{-4.5}-10^{-5} cm $

\hspace{5.4cm}(linear dimension of the particle of lower level)

According to this estimation, one space monad of the turbulent
field contains approximately $10^{9}-10^{10}$ nonstandard points
(the particles in the lower level). If the fluid is gas, there are
approximately $10^{5}-10^{6}$ gas molecules in one nonstandard
point of a monad field. This number of gas molecules is enough for
stable mean values in the statistical average.

\begin{quote}
     \it{Assumption 3: Turbulent field is continuous}.
\end{quote}

Here the meanings of continuity are: firstly the global turbulent
field is composed continuously of standard points (i.e., monad
fields); secondly the flows in the monad fields are continuous;
and thirdly the physical quantities on the interface between two
infinitely close monads are continuous as well. \noindent The
monads of $$\lim_{St\Delta x_{i}\rightarrow 0} (x_{i}+\Delta
x_{i},x_{j},x_{k},t)$$ and
$$(x_{i},x_{j},x_{k},t)$$ are
taken as two infinitely close monads in space. And so are in time.
The mathematic expression of the third is
\begin{equation}
  \lim_{St\Delta x\rightarrow 0}U(x+\Delta x,0)=U(x,L),
  \lim_{St\Delta t\rightarrow 0}U(t+\Delta t,0)=U(t,T)
\end{equation}

There is an even more
important meaning of the continuity. That will be stated in Assumption 6.

\section{Uncertainty Of Measurement}\indent
\indent To make a physical measurement, some time-space point
$(x_{1},x_{2}, x_{3},t)$ has to be assigned first, at which the
measurement is carried out. The results of the measurement are
taken as the measuring data of this point $(x_{1},x_{2},x_{3},t)$.
According to the standpoint of nonstandard analysis, the point
$(x_{1},x_{2},x_{3},t)$ is not an absolute point but a monad, in
which there are infinite nonstandard points. Generally speaking,
the motion features of different nonstandard points are different.
So the meaning of saying ``measuring at point
$(x_{1},x_{2},x_{3},t)$" becomes ambiguous. Obviously, if there
are fields in two different levels in the system being studied,
there must exist the new meaning of a measurement at any point of
the field in higher level.

\begin{quote}
     \it{Assumption 4: When a measurement at any point (monad) $(x_{1},x_{2},x_{3},t)$
     in a physical field is taken, the operation of the measurement
    will act randomly on one interior point (nonstandard point) of the point $(x_{1},x_{2},
    x_{3},t)$}.
\end{quote}

This assumption states clearly that one measurement operation at
some standard point (monad) in a field will always take place at
an interior point of such a monad. But the measurer can not
determine which interior point is acted on by the measurement
operation. What actually happens is random. The measurer knows
only the standard point (monad) in which the interior point acted
on by one measurement operation is contained. It is surprising
that when a measurement is carried out there is no way to
determine the exact position of the measured object. That is the
uncertainty of measurement. At first sight it seems that the
uncertainty of measurement stems from the rough measuring skill
and that the circumstance conditions of the measurement are not
exactly controllable etc.. After further thinking, it is
comprehended that the uncertainty rises essentially from limited
ability and means of the description of the physical world in
hierarchical structures. Apparently, in their quest of the laws of
the universe, people need not (and can not in fact) describe all
motion states of every molecule and electron in the universe.
Conversely, even though an exact and perfect description of the
motion states of every molecule and electron in the universe is
given, the laws of the motion of the whole universe are still not
understood. It is clearly that because the physical world has a
hierarchical structure, therefore any description and measurement
of physical phenomena should be related with the hierarchical
structure. When the effect of hierarchical structure can not be
ignored, the uncertainty of measurement occurs. The measurement
and observation of turbulence can be taken for examples of this
case.

An operation of measurement will act randomly on various interior
points in a monad. By what probability does the measurement
operation act on each interior point of a monad? In turbulence it
is thought that one operation of measurement will take place in
equiprobability at each interior point of a monad.

\begin{quote}
     \it{Assumption 5: When a measurement at any point (monad) of a turbulent
    field is made, the operation of the measurement will act in equiprobability on various
    interior points of the monad. This assumption is called the equiprobability
    assumption}.
\end{quote}

So there are two kinds of averages. One is that one measurement is
just taking average over a large number of molecules contained in
some interior point of a monad, while the other average over the
motions of all interior points in a monad has to be taken in order
to get the mean values of the physical quantities over the
standard point(monad) of the turbulent field. According to the
equiprobability assumption, the second average formula can be
given. Let $U(\vec{x},t,\vec{x^{\prime}} ,t^{\prime})$ represent
some physical quantity at an interior point $(\vec{x^{\prime}},
t^{\prime})$ of the monad $(\vec{x},t)$ and
$\widetilde{U}(\vec{x},t)$  express the average of
$U(\vec{x},t,\vec{x^{\prime}}, t^{\prime})$ over all interior
points of the monad. Then it follows that

    \begin{equation}
    \widetilde{U}(\vec{x},t)=\frac{1}{T}\int_{0}^{T}dt^{\prime}\frac{1}{L^{3}}\int
    _{0}^{L}dx_{1}^{\prime}\int_{0}^{L}dx_{2}^{\prime}\int_{0}^{L}dx_{3}^{\prime}
    U(\vec{x},t,\vec{x^{\prime}},t^{\prime})
    \end{equation}

Where the monad $(\vec{x},t)\equiv(x_{1},x_{2},x_{3},t)$, the
interior point of the monad
$(\vec{x^{\prime}},t^{\prime})\equiv(x_{1}^{\prime},x_{2}^{\prime},
x_{3}^{\prime},t^{\prime})$, and T, L are infinites of time and
space respectively. Let $T=L_{1}=L_{2}=L_{3}=L$ here and later.
Obviously, $\widetilde{U}$ is only a function of the standard
point $(\vec{x},t)$.

If the monad fields of a global flow field are uniform, there is
no difference between saying ``one measurement acting on the
interior point of a monad" and ``one measurement acting on the
point (monad)". Then the global field is laminar. Exactly,the
essential feature of turbulence is that its particle (standard
point) has internal structure, i.e., internal flow field (monad
field). That could be taken as the definition of turbulence.

Because of the uncertainty of measurement, one measurement at the
point $(\vec{x},t)$ will act randomly on any interior point.
Therefore, the results of many measurements carried out at the
same point(monad) under constant circumstance conditions will
indicate randomly the motion states of various interior points of
the monad. Surely fluctuation of data of the measurements must
appear, provided that there are considerable different states
among various interior points of the monad. This is the mechanism
of fluctuation appearing in a turbulent field.

Thus, a turbulent field is composed of monads with interior
structure (monad field). The flows in the monad fields are
controlled by the governing equations of (13)-(14). Still
continuity of turbulent field exists. Therefore, turbulence is
also regular flow. The appearance of the fluctuation of a
turbulent flow stems from the uncertainty of measurement and
observation in the turbulent field.

To keep the concepts mentioned above in mind, the order of
magnitude of fluctuation in a turbulent field can be estimated
roughly. Let $\Delta x^{\prime}$ be finite length in the
coordinate system of a monad, $U\sim 0(1)$ any physical quantity
and its fluctuation $u\sim U(x^{\prime}+\Delta
x^{\prime})-U(x^{\prime})\sim \frac{\partial U} {\partial
x^{\prime}}\Delta x^{\prime}$. The actual rate of change (i.e.,
observed from the angle of the global field level) of function U
is $\frac{\partial U}{\varepsilon^{2}
\partial x^{\prime}}$. In one case $\frac{\partial U}{\partial x^{\prime}}\sim
0(\varepsilon^{2})$, $u\sim 0(\varepsilon^{2})$, no obvious fluctuation occurs.
In another case $\frac{\partial U}{\partial x^{\prime}}\sim 0(1)$, and $\frac{\partial U}
{\varepsilon^{2}\partial x^{\prime}}\sim 0(\frac{1}{\varepsilon^{2}})$ showing
that there is excess rate of shear strain in the field. This will lead to
instability in the flow field. So the case should be excluded. Then in the third case
$\frac{\partial U}{\partial x^{\prime}}\sim 0(\varepsilon)$ only, and $\frac{
\partial U}{\varepsilon^{2}\partial x^{\prime}}\sim 0(\frac{1}{\varepsilon})$
with fluctuation $u\sim 0(\varepsilon)$. So there is notable
fluctuation in order of magnitude $u\sim 0(\varepsilon)$ in
turbulence provided that there exists any set of interior points
in a monad, and on the set $\frac{\partial U}{\partial x^{\prime}}
\sim 0(\varepsilon)$.

\section{Conservation Equations}\indent
\indent Generally speaking, in fluid mechanics a small volume
$\Delta x_{1}\Delta x_{2}\Delta x_{3}$ is taken and the flux of
conservative physical quantity through the boundary planes of the
volume is computed. Then let $\Delta x_{1}\rightarrow 0, \Delta
x_{2}\rightarrow 0, \Delta x_{3}\rightarrow 0$, the small volume
will tend to the point $(x_{1},x_{2},x_{3})$. Thus, conservation
equations at the point $(x_{1},x_{2},x_{3})$ are obtained. The
boundary planes of the small volume are absolute planes without
thickness. However, in nonstandard case a small volume is still
taken. By Use of the same method as in standard case but let
$St\Delta x_{1} \rightarrow 0, St\Delta x_{2}\rightarrow 0,
St\Delta x_{3}\rightarrow 0$, the small volume $\Delta x_{1}\Delta
x_{2}\Delta x_{3}$ will become the monad of $(x_{1},x_{2},x_{3})$.
Now the boundary planes of the small volume are not absolute
planes but very thin layers with thickness of $\varepsilon$ (the
linear dimension of the monad). Also conservation equations at
this monad are obtained. Surely the equations are those of mean
quantities over the monad in which infinite interior points are
contained.

Now these conservation equations are given in the following.

Let $``\sim"$ express the average operation over $(x_{1}^{\prime},
x_{2}^{\prime},x_{3}^{\prime},t^{\prime})$, $``\sim ^{i}"$ over $x_{i}^{\prime}$,
$``-"$ over $t^{\prime}$, etc., in a monad.

Through the pair of the planes perpendicular to $x_{1}$ axis the mass inflow
per unit time is:

 $$\widetilde{\widetilde{\overline{U_{1}}}^{1}}^{
\Delta x_{2}\Delta x_{3}}(x_{1})\Delta x_{2}\Delta x_{3}$$

\noindent the mass outflow

$$\widetilde{\widetilde{\overline{U_{1}}}^{1}}^{
\Delta x_{2}\Delta x_{3}}(x_{1}+\Delta x_{1})\Delta x_{2}\Delta x_{3}$$

\noindent the net inflow

 $$-\frac{\partial}{\partial x_{1}} \widetilde{\widetilde{\overline
    {U_{1}}}^{1}}^{\Delta x_{2}\Delta x_{3}}\Delta x _{1}\Delta x_{2}
    \Delta x_{3}$$

\noindent Through the other two pairs of the planes perpendicular
to the two other axes the net mass inflow per unit time are,
respectively,

    $$-\frac{\partial}{\partial x_{2}} \widetilde{\widetilde{\overline
    {U_{2}}}^{2}}^{\Delta x_{1}\Delta x_{3}}\Delta x _{1}\Delta x_{2}
    \Delta x_{3}$$

\noindent and

    $$-\frac{\partial}{\partial x_{3}} \widetilde{\widetilde{\overline
    {U_{3}}}^{3}}^{\Delta x_{1}\Delta x_{2}}\Delta x _{1}\Delta x_{2}
    \Delta x_{3}$$

\noindent Because of the law of mass conservation, it follows that

    $$\frac{\partial }{\partial x_{1}}\widetilde{\widetilde{\overline
    {U_{1}}}^{1}}^{\Delta x_{2}\Delta x_{3}}+\frac{\partial}{\partial x_{2}}
    \widetilde{\widetilde{\overline{U_{2}}}^{2}}^{\Delta x_{1}\Delta x_{3}}
    +\frac{\partial}{\partial x_{3}} \widetilde{\widetilde{\overline
    {U_{3}}}^{3}}^{\Delta x_{1}\Delta x_{2}}=0$$

\noindent Here the average of $``\sim \Delta x_{1}\Delta x_{2}"$
is taken over the plane of $\Delta x_{1}\Delta x_{2}$, which tends
to $``\sim 12"$ when $St\Delta x_{1} \rightarrow 0,St\Delta
x_{2}\rightarrow 0$. And so are the others. Let $St\Delta
x_{1}\rightarrow 0, St\Delta x_{2}\rightarrow 0, St\Delta
x_{3}\rightarrow 0$, then it is obtained that

    $$\frac{\partial }{\partial x_{1}}\widetilde{\overline
    {U_{1}}}^{123}+\frac{\partial }{\partial x_{2}}\widetilde{\overline
    {U_{2}}}^{123}+\frac{\partial }{\partial x_{3}}\widetilde{\overline
    {U_{3}}}^{123}=0$$

\noindent i.e.,

    \begin{equation}
    \frac{\partial\widetilde{U_{1}}}{\partial x_{1}}+\frac{\partial\widetilde{U_{2}}}
    {\partial x_{2}}+\frac{\partial\widetilde{U_{3}}}{\partial x_{3}}=0
    \end{equation}

Similarly, the equations of momentum conservation can be obtained
too.

Through the pair of the planes perpendicular to $x_{1}$ axis, the
net inflow of the momentum in $x_{1}$ direction per unit time is

    $$\rho\widetilde{\widetilde{\overline{U_{1}U_{1}}}^{1}}^{\Delta x_{2}
    \Delta x_{3}}(x_{1})\Delta x_{2}\Delta x_{3}-\  \rho\widetilde{\widetilde
    {\overline{U_{1}U_{1}}}^{1}}^{\Delta x_{2}\Delta x_{3}}(x_{1}+\Delta x_{1})
    \Delta x_{2}\Delta x_{3}$$ $$=-\rho\frac{\partial}{\partial x_{1}}\widetilde
    {\widetilde{\overline{U_{1}U_{1}}}^{1}}^{\Delta x_{2}
    \Delta x_{3}}\Delta x_{1}\Delta x_{2}\Delta x_{3}$$

\noindent Through the other two pairs of the planes perpendicular
to the other two axes, the net inflows of the momentum in $x_{1}$
direction per unit time are

    $$-\rho\frac{\partial}{\partial x_{2}}\widetilde{\widetilde{\overline
    {U_{2}U_{1}}}^{2}}^{\Delta x_{1}\Delta x_{3}}\Delta x_{1}\Delta x_{2}\Delta x_{3}$$

\noindent and

    $$-\rho\frac{\partial}{\partial x_{3}}\widetilde{\widetilde{\overline
    {U_{3}U_{1}}}^{3}}^{\Delta x_{1}\Delta x_{2}}\Delta x_{1}\Delta x_{2}\Delta x_{3}$$

\noindent respectively.

The force in $x_{1}$ direction exerting on the pair of the planes
perpendicular to $x_{1}$ axis is:

    $$-\widetilde{\widetilde{\overline{\sigma_{11}}}^{1}}^{\Delta x_{2}
    \Delta x_{3}}(x_{1})\Delta x_{2}\Delta x_{3}+\widetilde{\widetilde{\overline
    {\sigma_{11}}}^{1}}^{\Delta x_{2}\Delta x_{3}}(x_{1}+\Delta x_{1})
    \Delta x_{2}\Delta x_{3}$$ $$=\frac{\partial}{\partial x_{1}}\widetilde{\widetilde
    {\overline{\sigma_{11}}}^{1}}^{\Delta x_{2}
    \Delta x_{3}}\Delta x_{1}\Delta x_{2}\Delta x_{3}$$

\noindent The forces in $x_{1}$ direction exerting on
the other two pairs of the planes perpendicular to the other two axes are

    $$\frac{\partial}{\partial x_{2}}\widetilde{\widetilde{\overline
    {\sigma_{21}}}^{2}}^{\Delta x_{1}\Delta x_{3}}\Delta x_{1}\Delta x_{2}\Delta x_{3}$$

\noindent and

    $$\frac{\partial}{\partial x_{3}}\widetilde{\widetilde{\overline
    {\sigma_{31}}}^{3}}^{\Delta x_{1}\Delta x_{2}}\Delta x_{1}\Delta x_{2}\Delta x_{3}$$

\noindent respectively.

Then the increments of the momentum in $x_{1}$ direction in the small volume per unit time are

    $$\rho\{\widetilde{\overline{U_{1}}}^{\Delta x_{1}\Delta x_{2}\Delta x_{3}}
    (t+\Delta t)-\widetilde{\overline{U_{1}}}^{\Delta x_{1}\Delta x_{2}\Delta x_{3}}
    (t)\}\frac{\Delta x_{1}\Delta x_{2}\Delta x_{3}}{\Delta t}$$ $$=\rho\frac{\partial}
    {\partial t}\widetilde{\overline{U_{1}}}^{\Delta x_{1}\Delta x_{2}\Delta x_{3}}
    \Delta x_{1}\Delta x_{2}\Delta x_{3}$$

\noindent Because of the law of momentum conservation, it follows
that

    $$\rho\frac{\partial}
    {\partial t}\widetilde{\overline{U_{1}}}^{\Delta x_{1}\Delta x_{2}\Delta x_{3}}
    +\rho\frac{\partial}{
    \partial x_{1}}\widetilde{\widetilde{\overline{U_{1}U_{1}}}^{1}}^{\Delta x_{2}
    \Delta x_{3}}+\rho\frac{\partial}{\partial x_{2}}\widetilde{\widetilde{\overline
    {U_{2}U_{1}}}^{2}}^{\Delta x_{1}\Delta x_{3}}+\rho\frac{\partial}{\partial x_{3}}
    \widetilde{\widetilde{\overline{U_{3}U_{1}}}^{3}}^{\Delta x_{1}\Delta x_{2}}$$

    $$=\frac{\partial}{\partial x_{1}}\widetilde{\widetilde{\overline
    {\sigma_{11}}}^{1}}^{\Delta x_{2}\Delta x_{3}}+\frac{\partial}{\partial x_{2}}
    \widetilde{\widetilde{\overline{\sigma_{21}}}^{2}}^{\Delta x_{1}
    \Delta x_{3}}+\frac{\partial}{\partial x_{3}}\widetilde{\widetilde
    {\overline{\sigma_{31}}}^{3}}^{\Delta x_{1}\Delta x_{2}}$$

\noindent Let $St\Delta x_{1}\rightarrow 0, St\Delta x_{2}\rightarrow 0,
St\Delta x_{3}\rightarrow 0$, so $``\sim\Delta x_{1}\Delta x_{2}\Delta x_{3}"$
tend to $``\sim 123"$, etc. Then it follows that

    $$\frac{\partial}{\partial t}\widetilde{\overline{U_{1}}}^{123}+\frac{\partial}
    {\partial x_{1}}\widetilde{\widetilde{\overline{U_{1}U_{1}}}^{1}}^{23}+
    \frac{\partial}{\partial x_{2}}\widetilde{\widetilde{
    \overline{U_{2}U_{1}}}^{2}}^{13}+\frac{\partial}{\partial x_{3}}
    \widetilde{\widetilde{\overline{U_{3}U_{1}}}^{3}}^{12}$$
    $$=\frac{1}{\rho}\frac{\partial}{\partial x_{1}}\widetilde{\widetilde{
    \overline{\sigma_{11}}}^{1}}^{23}+\frac{1}{\rho}\frac{\partial}
    {\partial x_{2}}\widetilde{\widetilde{\overline{\sigma_{21}}}^{2}}^{13}
    +\frac{1}{\rho}\frac{\partial}{\partial x_{3}}\widetilde{\widetilde{
    \overline{\sigma_{31}}}^{3}}^{12}$$

\noindent i.e.,

    $$\frac{\partial}{\partial t}\widetilde{U_{1}}+\frac{\partial}
    {\partial x_{1}}\widetilde{U_{1}U_{1}}+
    \frac{\partial}{\partial x_{2}}\widetilde{U_{2}U_{1}}+\frac{\partial}
    {\partial x_{3}}\widetilde{U_{3}U_{1}}$$
    $$=\frac{1}{\rho}(\frac{\partial}{\partial x_{1}}\widetilde{\sigma_{11}}
    +\frac{\partial}{\partial x_{2}}\widetilde{\sigma_{21}}
    +\frac{\partial}{\partial x_{3}}\widetilde{\sigma_{31}})$$

Similarly, the momentum equations in $x_{2}$ and $x_{3}$
directions can be written too. The general form of momentum
equations is:

\begin{equation}
    \frac{\partial \widetilde{U_{i}}}{\partial t}+\frac{\partial \widetilde{U_{i}U_{j}}}
    {\partial x_{j}}=\frac{1}{\rho}\frac{\partial \widetilde{\sigma_{ij}}}{\partial x_{j}}
\end{equation}

The equations $(17)-(18)$ are about the mean quantities over a
monad and called the conservation equations. Here no special
mean-methods are assigned. But the mean $``\sim"$ will always be
taken
 as the mean-method showed in (16) in this paper.

\section{Fundamental Equations of Turbulent Flows }\indent
\indent Still instantaneous quantities $U_{i}$ and $P$ are
decomposed into two parts such that
\begin{equation}
    U_{i}=\widetilde{U_{i}}+u_{i},\quad P=\widetilde{P}+p ,\qquad \widetilde{u_{i}}=0,\quad\widetilde{p}=0
\end{equation}

Here the quantities with  $``\sim"$ are called mean quantities; in
this regard, we will adopt the mean-method of (16) i.e.,
\begin{equation}
    \widetilde{U}=\frac{1}{T}\int_{0}^{T}dt^{\prime}\frac{1}{L^{3}}
    \int_{0}^{L}dx_{1}^{\prime}\int_{0}^{L}dx_{2}^{\prime}\int_{0}^{L}dx_{3}^{\prime}U,\quad
    \overline{U}=\frac{1}{T}\int_{0}^{T}dt^{\prime}U,\quad\widetilde{U}^{i}=
    \frac{1}{L}\int_{0}^{L}dx_{i}^{\prime}U
\end{equation}
    $$\widetilde{U}^{ij}=
    \frac{1}{L^{2}}\int_{0}^{L}dx_{i}^{\prime}\int_{0}^{L}dx_{j}^{\prime}U,
    \quad\widetilde{U}^{ijk}=\frac{1}{L^{3}}\int_{0}^{L}dx_{i}^{\prime}
    \int_{0}^{L}dx_{j}^{\prime}\int_{0}^{L}dx_{k}^{\prime}U$$

\indent The relations of (19) are the definitions of $u_{i}$ and
$p$. The $u_{i}$ and $p$ have the same the order of magnitude as
the turbulent fluctuation mentioned in Section 3. We will call the
$u_{i}$ and $p$ in (19) as fluctuant quantities. To note that the
$\widetilde{U_{i}}$ and $\widetilde{P}$ are mean quantities over a
monad, and $u_{i}$ and $p$ are meaningful to a monad (or in a
monad). They have different meanings with those in the existent
theory of turbulence.

\indent Before derivation of the fundamental equations of
turbulent flows, it is necessary to present the assumption of
close property between two monads when they are infinitely close
to each other, and analyze the mutual relation of the two monads
in motion characteristics.

\begin{quote}
  \it{Assumption 6: In both the value and structure
  of function, physical function, defined on the interior points of the monads
  of a turbulent field, is very close between two monads,  when these two monads
  are infinitely close to each other}.
\end{quote}

Considering any physical quantity $U_{i}\sim 0(1)$, its
fluctuation $u_{i}$, the mathematical expressions of this
assumption are as follows:

\begin{equation}
     (a)\ \delta u_{i}\sim 0(\varepsilon^{2}), \qquad
     (b)\ \delta(u_{i}u_{j}-\widetilde{u_{i}u_{j}})\sim 0(\varepsilon^{4})
\end{equation}
      $$(0\leq x_{i}^{\prime}<L,\quad 0\leq t^{\prime}<T)$$

\noindent Here the operator $\delta$ is defined as
\begin{equation}
   \delta_{i} U=\lim_{St\Delta x_{i}\rightarrow 0}[U(x_{i}+\Delta x_{i},x_{i}^{\prime})
   -U(x_{i},x_{i}^{\prime})],\qquad
   \delta_{t} U=\lim_{St\Delta t\rightarrow 0}[U(t+\Delta t,t^{\prime})-U(t,t^{\prime})]
\end{equation}

It can be got from (a) in (21) that
 \begin{equation}
     \ \ \delta \left[\frac{\partial U}{\varepsilon^{2}\partial x_{i}^{\prime}}-
     \frac{\partial U}{\partial x_{i}}\right]\sim 0(\varepsilon^{2})
 \end{equation}
Where $\frac{\partial U}{\varepsilon^{2}\partial x_{i}^{\prime}}$
is the actual rate of change or figure slope of the function $U$
at the nonstandard point $x_{i}^{\prime}$, and $\frac{\partial
U}{\partial x_{i}}$ is the mean value of $\frac {\partial
U}{\varepsilon^{2}\partial x_{i}^{\prime}}$ over $x^{\prime}_{i}$
(see(25)). Clearly the condition (23) and $ (a)$ in (21) show that
the figure shape of the function $U$ has very little difference
between two infinitely close monads. Similarly the condition $(b)$
in (21) shows that the figure shape of the function $u_{i}u_{j}$
has very little difference between two infinitely close monads.

Assumption 6 could be called ``Continuous Assumption". It is well
known that with a few exceptions, physical quantities  vary
usually continuously with time and space. In the case of standard
analysis, a function is defined on the standard points. It is
characterized only by its value. The continuity of a physical
quantity only requires that the function representing the physical
quantity be continuous in mathematics. On the other hand, in the
nonstandard case a point becomes a monad and a function is defined
at the interior point of the monad. There exists a
function-structure on the whole monad. So the function is
characterized by not only its value but also its structure on the
monad. The continuity of a physical function will mean that in
both the values and figure-shape of the function , the variation
of the physical function with monads is continuous. Thus, there is
very little difference between two infinitely close monads in both
value and structure of the physical function. In other words, the
Assumption 6 is the nonstandard expansion of the physical function
continuity in the case of standard analysis. Moreover, the
Assumption 6 is reasonable and natural, because the circumstance
conditions imposed on two infinitely close monads are very close
to each other.

  In a monad, we consider that stress tensor for Newtonian
fluid is
\begin{equation}
    \sigma_{ij}=-P\delta_{ij}+\mu\left(\frac{\partial U_{i}}{\partial x_{j}^{\prime}}
    +\frac{\partial U_{j}}{\partial x_{i}^{\prime}}\right)\frac{1}{\varepsilon^{2}}
\end{equation}

\noindent Using (20) and(17),(15) (Assumption 3) we can give the
following computation:
\begin{equation}
   \frac{1}{\varepsilon^{2}}\widetilde{\frac{\partial U_{i}}{\partial x_{j}
    ^{\prime}}}^{j}=\left(\frac{\partial U_{i}}{\partial x_{j}}\right)_{x^{\prime}_{j}=0}
\end{equation}

\noindent Then
$$\widetilde{\sigma_{ij}}=-\widetilde{P}\delta_{ij}+\mu{\left
(\frac{\partial \widetilde{\overline {U_{i}}}^{kl}}{\partial x_{j}}\right)_{x^{\prime}_{j}=0}}
 +\mu{\left
(\frac{\partial \widetilde{\overline {U_{j}}}^{mn}}{\partial x_{i}}\right)_{x^{\prime}_{i}=0}}$$

\begin{eqnarray*}
\frac{\partial\widetilde{\sigma_{ij}}}{\partial x_{j}}&=& -\frac{\partial\widetilde{P}}
{\partial x_{i}}+\mu\frac{\partial}{\partial x_{j}}\left
(\frac{\partial \widetilde{\overline {U_{i}}}^{kl}}{\partial x_{j}}\right)_{x^{\prime}_{j}=0}
+\mu\frac{\partial}{\partial x_{j}}\left
(\frac{\partial \widetilde{\overline {U_{j}}}^{mn}}{\partial x_{i}}\right)_{x^{\prime}_{i}=0}\\
&=& -\frac{\partial\widetilde{P}}
{\partial x_{i}}+\mu\frac{\partial}{\partial x_{j}}\frac{\partial\widetilde{U_{i}}}{\partial x_{j}}
+\mu\frac{\partial}{\partial x_{j}}\left
(\frac{\partial \widetilde{\overline {u_{i}}}^{kl}}{\partial x_{j}}\right)_{x^{\prime}_{j}=0}
\end{eqnarray*}

\noindent By use of (a) in (21) it follows that
\begin{equation}
  \left(\frac{\partial^{2}\widetilde{\overline{u_{i}}}^{kl}}
  {\partial x^{2}_{j}}\right)_{x^{\prime}_{j}=0}\sim 0(\varepsilon)
\end{equation}
To omit them, it is obtained that
\begin{equation}
   \frac{\partial\widetilde{\sigma_{ij}}}{\partial x_{j}}=-\frac{\partial\widetilde{P}}
   {\partial x_{i}}+\mu\frac{\partial^{2}\widetilde{U_{i}}}{\partial x^{2}_{j}}
\end{equation}

\noindent Then the equations (18) can be written as
\begin{equation}
    \frac{\partial \widetilde{U_{i}}}{\partial t}+\frac{\partial \widetilde{U_{i}U_{j}}}
    {\partial x_{j}}=-\frac{1}{\rho}\frac{\partial \widetilde{P}}{x_{i}}+\nu \nabla^{2}
    \widetilde{U_{i}},\qquad \nabla^{2}=\frac{\partial}{\partial x_{j}}\frac{\partial}
    {\partial x_{j}}
\end{equation}
\noindent(17) is written as
\begin{equation}
   \frac{\partial \widetilde{U_{i}}}{\partial x_{i}}=0
\end{equation}
\noindent Here $\nu=\frac{\mu}{\rho}$ the kinematic viscosity.

Note that now the mean method in the equations (28) and (29) is
defined by (20). Then by use of the relation (19), the
decomposition of (29) and (28) can be done as

\begin{equation}
    \frac{\partial U_{i}}{\partial x_{i}}-\frac{\partial u_{i}}{\partial x_{i}}=0
\end{equation}

\begin{equation}
    \frac{\partial U_{i}}{\partial t}+\frac{\partial U_{i}U_{j}}{\partial x_{j}}-\frac
    {\partial u_{i}}{\partial t}-\frac{\partial (U_{i}U_{j})_{fl}}{\partial x_{j}}$$  $$=
    -\frac{1}{\rho}\frac{\partial P}{\partial x_{i}}+\frac{1}{\rho}\frac{\partial p}
    {\partial x_{i}}+\nu\nabla^{2}U_{i}-\nu\nabla^{2}u_{i}
\end{equation}

\noindent Here $u_{i},p$ and $(U_{i}U_{j})_{fl}$ are the
fluctuations of the velocity, pressure and $(U_{i}U_{j})$
respectively. The independent variables of the functions in the
equations (30)-(31) are
$(x_{1},x_{2},x_{3},t,x_{1}^{\prime},x_{2}^{\prime},
x_{3}^{\prime},t^{\prime})$. Based on Assumption 6 the order of
magnitude of the terms concerned with fluctuation quantities is
$\sim0(\varepsilon)$, but the order of magnitude of the terms
concerned with instantaneous quantities is $\sim0(1)$. So when
(30)-(31) are split into two parts having different order of
magnitude, the following can be obtained:

Equations, in which terms are in the order of magnitude $\sim0(1)$, of instantaneous
quantities are:
\begin{equation}
    \frac{\partial U_{i}}{\partial x_{i}}=0
\end{equation}

\begin{equation}
     \frac{\partial U_{i}}{\partial t}+\frac{\partial U_{i}U_{j}}{\partial x_{j}}=-\frac{1}
     {\rho}\frac{\partial P}{\partial x_{i}}+\nu\nabla^{2}U_{i}
\end{equation}

Equations, in which terms are in the order of magnitude $\sim0(\varepsilon)$, of fluctuation
quantities are:
\begin{equation}
    \frac{\partial u_{i}}{\partial x_{i}}=0
\end{equation}

\begin{equation}
    \frac{\partial u_{i}}{\partial t}+\frac{\partial (U_{i}U_{j})_{fl}}{\partial x_{j}}
    =-\frac{1}{\rho}\frac{\partial p}{\partial x_{i}}+\nu\nabla^{2}u_{i}
\end{equation}

\noindent Where $(U_{i}U_{j})_{fl}$ in (35) is:

\begin{eqnarray*}
    (U_{i}U_{j})_{fl}&=&U_{i}U_{j}-\widetilde{U_{i}U_{j}}\\
                     &=&\widetilde{U_{i}}u_{j}+\widetilde{U_{j}}u_{i}+u_{i}u_{j}
                        -\widetilde{u_{i}u_{j}}
\end{eqnarray*}

\noindent Substitution of the relations into (35) produces
\begin{equation}
    \frac{\partial u_{i}}{\partial t}+u_{j}\frac{\partial \widetilde{U_{i}}}{\partial x_{j}}
    +\widetilde{U_{j}}\frac{\partial u_{i}}{\partial x_{j}}+\frac{\partial u_{i}u_{j}}
    {\partial x_{j}}-\frac{\partial \widetilde{u_{i}u_{j}}}{\partial x_{j}}=-\frac{1}{\rho}
    \frac{\partial p}{\partial x_{i}} +\nu\nabla^{2}u_{i}
\end{equation}

Equations (28)(29),(32)(33) and (34)(35) are the governing
equations about mean, instantaneous and fluctuant quantities
respectively. They can be called as the fundamental equations of
turbulent flows.

It is found easily that these equations have the same form as
those in the existent theory of turbulent flows. But there are
essentially differences between the two. These differences are:

1.Different methods to get them. In the existent theory of
turbulence the mean equations are obtained by the average of the
instantaneous equations (i.e., the Navier-Stokes equations). Then
the mean equations are subtracted from the instantaneous equations
to get the fluctuation equations. But in this paper, the mean
equations will FIRST be obtained in view of the physical
conservation laws, and then by virtue of the close property
assumption, the instantaneous and fluctuant equations are obtained
respectively.

2.The terms in the equations of the existent theory are the
limits, such as
$$\lim_{\triangle x\rightarrow 0}\frac{f(x+\triangle x)-f(x)}{\triangle
x} \quad\quad \lim_{\triangle t\rightarrow 0}\frac{f(t+\triangle
t)-f(t)}{\triangle t}$$

\noindent But the terms in the equations of the nonstandard theory
are the partial derivatives defined in (10). The equations in two
theories are conceptually different. The former is based on, so
called, the frame of $\delta-\varepsilon$, the latter is out of
the frame of $\delta-\varepsilon$. Therefore, the equation in
nonstandard picture of turbulence is the new kind of equation.

3. The physical quantities in the equations of the existent
theory, such as velocity, pressure etc., are all functions of the
standard point $(\vec{x},t)$.  However, those in the nonstandard
case are all functions of the point
$(\vec{x},t,\vec{x^{\prime}},t^{\prime})$, actually functions of
the nonstandard point $(\vec{x^{\prime}},t^{\prime})$ with
$(\vec{x},t)$ showing which monad contains it. It is reasonable
that the physical quantities are taken as the functions of the
standard point $(\vec{x},t)$ in laminar flow, but not in turbulent
flows, because a standard point becomes a monad in which there is
an interior structure (i.e., the monad flow field) in the case of
turbulence. In fact, there is no definition of instantaneous and
fluctuant quantities of a whole monad, but there is the definition
of instantaneous and fluctuant quantities of the interior points
of the monad. So for the global field,the Navier-Stokes equations
in existent meanings do not exist in principle. Therefore, though
(32)-(33) are the same as the Navier-Stokes equations in forms,
they are different. The fundamental equations obtained in this
paper are governing equations about the physical quantities
defined at those interior points of various monads. These interior
points have the same nonstandard coordinates
$(\vec{x^{\prime}},t^{\prime})$. But the fundamental equations in
the existent turbulent theory are governing equations about the
physical quantities defined at the standard points of the global
flow field. There is an essential difference between them.

4. The average in the existent theory usually means  the ensemble
average, but the distribution function can not be given. So the
closure problem continues to exist. Otherwise, it is definite in
the nonstandard model to take average over all interior points of
a monad by the equiprobability assumption.

5. It should be pointed out that the function values at some
standard points are given for the initial and boundary conditions
of the fundamental equations in the existent theory. However, only
the function values at some nonstandard points can be taken as the
initial and boundary conditions for the instantaneous and
fluctuation quantities in the fundamental equations in the
nonstandard case. Because a turbulent field always starts and
develops from walls or laminar flows and evolves from laminar
flows or static fluids, the function values on walls or laminar
flows can be taken as the boundary conditions, and the conditions
of static fluids or laminar flows can be taken as the initial
conditions. No slip condition is still kept. Thus, the initial and
boundary conditions of the fundamental equations obtained from the
nonstandard analysis can be given in principle.

\section{On Closure Problem}\indent
\indent Substitution of the relation
$U_{i}=\widetilde{U_{i}}+u_{i}$ into (28) produces

\begin{equation}
    \frac{\partial \widetilde{U_{i}}}{\partial t}+\frac{\partial \widetilde{U_{i}}
    \widetilde{U_{j}}}{\partial x_{j}}+\frac{\partial \widetilde{u_{i}u_{j}}}{\partial x_{j}}
    =-\frac{1}{\rho}\frac{\partial \widetilde{P}}{x_{i}}+\nu \nabla^{2}
    \widetilde{U_{i}}
\end{equation}

\noindent This is the same in form as famous Reynolds equation. It
is not closure, and neither is equation (36) in which there is
term $\widetilde{u_{i}u_{j}}$. The closure problem stemming from
nonlinearity is very difficult in turbulence research. Now in the
nonstandard theory, the turbulence is composed of monad fields,
and the point-average is adopted. Then by the close property
between two infinitely close monads, we can give the reasonable
closure methods as follows.

By the relation (21)(Assumption 6) and the order of magnitude of
fluctuation $u_{i}\sim 0(\varepsilon)$, it follows that

    $$\delta(u_{i}u_{j})=u_{i}\delta u_{j}+u_{j}\delta u_{i}\sim0(\varepsilon^{3})$$
    $$\delta \widetilde{u_{i}u_{j}}\sim 0(\varepsilon^{3})$$\quad\
    $$\frac{\partial \widetilde{u_{i}u_{j}}}{\partial x_{j}}\sim 0(\varepsilon^{2})$$
\noindent and
    $$\delta(u_{i}u_{j}-\widetilde{u_{i}u_{j}})\sim 0(\varepsilon^{4})$$
\noindent also,it follows that
\begin{equation}
    \frac{\partial (u_{i}u_{j}-\widetilde{u_{i}u_{j}})}{\partial x_{j}}\sim 0(\varepsilon^{3})
\end{equation}
  $${}$$
\indent There are three choices for the closure of equations:

\noindent (A) Choice one:

\indent The term $\frac{\partial \widetilde{u_{i}u_{j}}}{\partial
x_{j}}\sim 0(\varepsilon^{2})$ is neglected from equations (37).
The closed equations of the mean quantities will be obtained as
follows.
\begin{equation}
   \frac{\partial \widetilde{U_{i}}}{\partial x_{i}}= 0
\end{equation}
\begin{equation}
   \frac{\partial\widetilde{U_{i}}}{\partial t}+\frac{\partial \widetilde{U_{i}}\widetilde{U_{j}}}
   {\partial x_{j}}=-\frac{1}{\rho}\frac{\partial \widetilde{P}}{\partial
   x_{i}}+\nu\nabla^{2}\widetilde{U_{i}}
\end{equation}
$${}$$
\noindent (B) Choice two:

\indent To note that

    \quad\quad $$\frac{\partial (u_{i}u_{j}-\widetilde{u_{i}u_{j}})}{\partial x_{j}}\sim
    0(\varepsilon^{3})\quad\quad\quad\quad\quad \quad\quad\quad (38)$$

\noindent and the other terms in equation (36) have the order of
magnitude of $\sim 0(\varepsilon)$. Hence, the term (38) can be
neglected from (36). Then the equation (36) becomes
\begin{equation}
    \frac{\partial u_{i}}{\partial t}+u_{j}\frac{\partial \widetilde{U_{i}}}{\partial x_{j}}
    +\widetilde{U_{j}}\frac{\partial u_{i}}{\partial x_{j}}=-\frac{1}{\rho}
    \frac{\partial p}{\partial x_{i}} +\nu\nabla^{2}u_{i}
\end{equation}

\noindent With $\widetilde{U_{i}}=U_{i}-u_{i}$, this equation
changes to
\begin{equation}
    \frac{\partial u_{i}}{\partial t}+U_{j}\frac{\partial u_{i}}{\partial x_{j}}
    +u_{j}\frac{\partial U_{i}}{\partial x_{j}}-2u_{j}\frac{\partial u_{i}}{\partial x_{j}}
    =-\frac{1}{\rho}\frac{\partial p}{\partial x_{i}} +\nu\nabla^{2}u_{i}
\end{equation}

Under certain initial and boundary conditions, the equations
(32)-(33) are solved and $U_{i}$ are obtained. Then the equations
(42) with (34) becomes also closed, the fluctuations $u_{i}$ and
$p$ can be obtained after the equations (34)(42) are solved.
Finally the mean quantities $\widetilde{U_{i}}=U_{i}-u_{i}$ and
$\widetilde{P}=P-p$.
$${}$$
\noindent (C) Choice three:

\indent Hence, the term (38) can be neglected from (36) and (37).
Then it follows that
\begin{equation}
\frac{\partial \widetilde{U_{i}}}{\partial x_{i}}=0
\end{equation}

\begin{equation}
\frac{\partial \widetilde{U_{i}}}{\partial t}+\frac{\partial
\widetilde{U_{i}}\widetilde{U_{j}}}{\partial x_{j}}+\frac{\partial
u_{i}u_{j}}{\partial x_{j}}=-\frac{1}{\rho}\frac{\partial
\widetilde{P}}{\partial
x_{i}}+\nu\nabla^{2}\widetilde{U_{i}}
\end{equation}

\begin{equation}
\frac{\partial u_{i}}{\partial x_{i}}=0
\end{equation}

\begin{equation}
\frac{\partial u_{i}}{\partial t}+\widetilde{U_{j}}\frac{\partial
u_{i}}{\partial x_{j}}+u_{j}\frac{\partial
\widetilde{U_{i}}}{\partial x_{j}}=-\frac{1}{\rho}\frac{\partial
p}{\partial x_{i}}+\nu\nabla^{2}u_{i}
\end{equation}

\indent Obviously, the equations (43)-(46) are closed. We can
obtain $\widetilde{U_{i}},\widetilde{P},u_{i},p,$ from solving
these equations.

Surely if $u_{i}=0, p=0$, i.e., no fluctuations occur, the
equations (28)(29) and (32)(33) are reduced to ordinary
Navier-Stokes equations and the field becomes a laminar field.

\section{Some Remarks on the Numerical Calculations}\indent
\indent To take the $\frac{\partial U}{\partial t}$ for example,
in the standard case we have
   $$\left(\frac{\partial U}{\partial t}\right)_{st}=\lim_{\Delta t\rightarrow 0}
   \frac{U(x_{1},x_{2},x_{3},t+\Delta t)-U(x_{1},x_{2},x_{3},t)}{\Delta t} $$
but in the nonstandard case by the definition of
$\left(\frac{\partial U}{\partial t}\right)_{nst}$ in this paper,
it follows that
\begin{equation}
  \left(\frac{\partial U}{\partial t}\right)_{nst}=\lim_{St\Delta t\rightarrow 0}
  \frac{U(x_{1},x_{2},x_{3},t+\Delta t,x^{\prime}_{1},x^{\prime}_{2},x^{\prime}_{3},t^{\prime})
  -U(x_{1},x_{2},x_{3},t,x^{\prime}_{1},x^{\prime}_{2},x^{\prime}_{3},t^{\prime})}{\Delta t}
\end{equation}

In form, there is no difference between the two cases when the
discretization of time-space is taken for the numerical
computation. However within a grid, obtained from the
discretization of $\left(\frac{\partial U}{\partial
t}\right)_{st}$, there is not structure (i.e., the whole grid
should be uniform). But the interior structure is permitted
through the grid obtained from the discretization of
$\left(\frac{\partial U}{\partial t}\right)_{nst}$. This diference
between two cases is very important.

Surely in the nonstandard case, $U(x_{1},x_{2},x_{3},t+\Delta
t,x^{\prime}_{1},x^{\prime}_{2}, x^{\prime}_{3},t^{\prime})$ and
$U(x_{1},x_{2},x_{3},t,x^{\prime}_{1},x^{\prime}_{2},
x^{\prime}_{3},t^{\prime})$ are the values of physical quantity
$U$ at the interior point
$(x^{\prime}_{1},x^{\prime}_{2},x^{\prime}_{3},t^{\prime})$ of two
different monads in time, that is monad of $(t+\Delta t)$ and
monad of $(t)$. Though the fundamental equations work for any
interior point
$(x^{\prime}_{1},x^{\prime}_{2},x^{\prime}_{3},t^{\prime})$ of
every monad, it is better that let $x^{\prime}_{1}=x^{\prime}_{2}=
x^{\prime}_{3}=t^{\prime}=0$ because the point
$(x_{1},x_{2},x_{3},t,0,0,0,0)$ is located on the boundary of the
time-space of a turbulent field, when $(x_{1},x_{2},x_{3},t)$ is
the boundary monad of the field.

Moreover, $U(x_{1},x_{2},x_{3},t+\Delta t,0,0,0,0)$ and
$U(x_{1},x_{2},x_{3},t,0,0,0,0)$ are hyperreal numbers. In the
numerical calculations, their standard parts will be taken.

Therefore, after the discretization, $\frac{\partial U}{\partial
t}$ becomes:
$$\frac{StU(x_{1},x_{2},x_{3},t+\Delta t,0,0,0,0)
  -StU(x_{1},x_{2},x_{3},t,0,0,0,0)}{\Delta t}$$

\noindent For simplicity it can also be written as

\begin{equation}
    \frac{U(x_{1},x_{2},x_{3},t+\Delta t,0,0,0,0)
    -U(x_{1},x_{2},x_{3},t,0,0,0,0)}{\Delta t}
\end{equation}
\noindent provided it is kept in mind that the $U$ in (48)
represents the standard parts of them.

There are similar discussions for the space partial derivertives of physical quantities.

After the $U(x_{1},x_{2},x_{3},t,0,0,0,0)$ and $u(x_{1},x_{2},x_{3},t,0,0,0,0)$,
the instantaneous and fluctuant quantities respectively, are obtained, then
\begin{equation}
 \widetilde{U}(x_{1},x_{2},x_{3},t)=U(x_{1},x_{2},x_{3},t,0,0,0,0)-u(x_{1},x_{2},x_{3},t,0,0,0,0)
\end{equation}
\noindent That is the mean value over the small volume of $\Delta
x_{1}\Delta x_{2}\Delta x_{3}\Delta t$, but it is taken as the
mean value over the point $(x_{1},x_{2},x_{3},t)$.

Moreover, the mean value $\widetilde{U}$ obtained by the methods
mentioned above, is the mean over the point(monad) and  still
random oscillatory. The random oscillation of the mean values
$\widetilde{U}$ ought to stem from the unavoidable random
disturbances in real turbulent fields. If the average of these
mean values over a finite time period(or a finite space range) is
taken once again, the final average results could be compared with
the measured average values of physical quantities over
corresponding time period(or space range).

\section{Conclusions}\indent
\indent In a view of the concepts of the nonstandard analysis and
the fact that the physical world has a hierarchical structure, one
nonstandard picture on turbulent flows has been presented in this
paper. The key points of this picture are as follows:

1. There exist two kinds of fields in different levels. One is the
global turbulent field composed of the standard points (monads),
and the other is the monad field. And as a whole a monad is not
already uniform. Now the interior structure occurs in a monad;
namely, different interior points of the monad possess different
characteristics of motion.

2. The flows in a monad field are governed by the Navier-Stokes
equations.

3. If there are two kinds of fields in different levels, one
operation of measurement at any point (actually a monad) of the
field in a higher level will act randomly on one interior point of
the monad field. The measurer can determine the monad, rather than
its interior point on which the measurement operation acts.

The position of measured object can not be determined exactly, but
over a range. Evidently there exists some uncertainty of
measurement. That uncertainty stems from the description and
observation of the physical phenomena in a hierarchical structure.

4. The flows in a turbulent field are also continuous and regular,
while they are seen intuitively in disorderly and irregular form.
Fundamentally, the disorder, irregularity and fluctuation occur
because of the uncertainty of measurement. Hence, we can probably
say that the disorder of turbulence stems from two sources: the
first is the fluctuation because of the uncertainty of
measurement. This kind of fluctuation is thought to be ``real
turbulent fluctuation". The second is the violent unsteadiness and
unhomogeneity of the field. The values of $\widetilde{U_{i}}$ and
$\widetilde{P}$ vary fast with time and space. It is due to the
flow-instability as the Reynolds number is very high. In this
case, the small random disturbances will be amplified.

5. There are two kinds of averages. The values of the measurement
acting on one interior point of a monad, merely stand for the mean
results over a lot of fluid molecules contained in this interior
point. The other is the average over all interior points in a
monad. The second average results indicate the mean values of the
physical quantities over the monad (the standard point) in the
global field. By the equiprobability assumption, the mean formula
of the second can be given in (20).

6. As the expansion of the fact that the physical functions are
continuous in the standard case, the assumption of close property
between two infinitely close monads shows that the variation of
interior structure as well as values of the physical function
 $U(\vec{x},t,\vec{x^{\prime}},t^{\prime})$ with $(\vec{x},t)$ (monad) is
continuous too. In other words, there exists only very little
difference in characteristics of the function
$U(\vec{x},t,\vec{x^{\prime}},t^{\prime})$ between two infinitely
close monads. According to this assumption and the conservation
laws, the governing equations of mean, instantaneous and fluctuant
quantities in turbulence (i.e., the fundamental equations) are
obtained, and the reasonable closure methods are given too. It
should be noted that the point-average is used, instead of the
Reynolds average, therefore the closure problem is easily overcome
in the nonstandard case.

7.The new kind of equations of turbulent motion is presented in
this paper. These equations are out of the frame of
$\delta-\varepsilon$ and valid for the points, which are permitted
to be not uniform. These points, in fact, are monads.

  $${}$$

\end{document}